# Realizing tunable Fermi level in SnTe by defect control

**Bamidele Onipede[1], Matthew Metcalf[1], Nisha Fletcher[3], Hui Cai[1,2,*]**

[1]Department of Physics, University of California, Merced, CA, USA
[2]Materials Sciences Division, Lawrence Berkeley National Laboratory, Berkeley, CA, USA
[3]Department of Mathematics, Southern Utah University, Cedar City, UT, USA.

**Email**:
Hcai6@ucmerced.edu

## Abstract

The tuning of the Fermi level in tin telluride, a topological crystalline insulator, is essential for accessing its unique surface states and optimizing its electronic properties for applications such as spintronics and quantum computing. In this study, we demonstrate that the Fermi level in tin telluride can be effectively modulated by controlling the tin concentration during chemical vapor deposition synthesis. By introducing tin-rich conditions, we observed a blue shift in the X-ray photoelectron spectroscopy core-level peaks of both tin and tellurium, indicating an upward shift in the Fermi level. This shift is corroborated by a decrease in work function values measured via ultraviolet photoelectron spectroscopy, confirming the suppression of Sn vacancies. Our findings provide a low-cost, scalable method to achieve tunable Fermi levels in tin telluride, offering a significant advancement in the development of materials with tailored electronic properties for next-generation technological applications.

Keywords: Tin telluride, Topological crystalline insulator, Fermi level, Defect engineering, Work function, X-ray photoelectron Spectroscopy (XPS), Ultraviolet photoelectron spectroscopy (UPS).

---

## 1. INTRODUCTION

Topological crystalline insulators (TCIs) are a class of materials that exhibit insulating behaviour in their bulk but possess conducting states on their surfaces or edges due to the underlying crystalline symmetry[1–3]. Unlike topological insulators (TIs), whose surface states are protected by time-reversal symmetry, the surface states in TCIs are protected by the symmetries of the crystal lattice, such as mirror symmetry, rotational symmetry, or other point group symmetries[4–10]. Among the various materials identified as TCIs, tin telluride (SnTe) has emerged as a prominent example due to its relatively simple crystal structure. SnTe crystallizes in a rock-salt structure with the space group $Fm\bar{3}m$. The nontrivial topology emerges from the mirror symmetry with respect to the (110) plane. As a result, SnTe possesses topological surface states as Dirac cones on a variety of crystal surfaces including the {100},{110} and {111} high symmetry surfaces of face-centered cubic structure[2]. The unique electronic properties of SnTe as a TCI open up exciting possibilities for various technological applications. The robustness of the surface states against perturbations makes SnTe a promising candidate for spintronic devices, where the spin of electrons is utilized for information processing. Additionally, the potential for realizing Majorana fermions in superconducting SnTe thin films could pave the way for advancements in quantum computing, particularly in the development of fault-tolerant qubits[10–12].

Tuning of the Fermi level ($E_F$) is very critical in topological insulators (TIs) in order to access the topological surface states. The desirable configuration for certain applications is that the Fermi level crosses the Dirac point of the topological insulator[7,13], such that the topological surface states dominate the bulk states. In the presence of defects, the Fermi level shifts into the conduction or valence band leading to available states that bulk carriers can occupy. Hence, the bulk carrier contribution would overshadow those from the topological surface states, making it difficult to observe and utilize the exotic properties of the topological surface states[4,6,7,13–16]. For example, a large bulk resistivity is required to observe surface quantum oscillations in TIs[17]. Additionally, the quantum anomalous Hall states which leads to dissipation free transport can only be observed when the small surface bandgap is hosted with high precision at the Fermi level[7]. Various methods have been studied to tune the position of the Fermi level including defect engineering, doping, etc.[18–22].

A characteristic defect exhibited by SnTe are tin (Sn) vacancies which act like acceptors and result in p-type doping in SnTe as it hosts intrinsic high-density holes in the bulk[4,18,23,24]. The difficulty in obtaining quantum transport measurements of the unique surface states in SnTe has been attributed to these intrinsic Sn vacancies[6,25] as the conduction from the bulk dominates the conduction from the surface states. Various approaches have been taken to suppress the intrinsic Sn-vacancies in SnTe[26–30], such as co-doping with Ag and Cu[31], doping with Bi[32]l, Sb[33], and Pb[34] and transition metal Mn[35]. Prominent among these approaches is substituting Sn with Pb and forming $Pb_xSn_{1-x}Te$ alloys[34,36]. However, this approach leads to a reduction of the bulk band gap and the alloy tends to form a topological trivial insulator phase above a certain concentration of Pb[3,34,37]. Therefore, it is demanding to find an efficient method to control Sn vacancies and achieve a tunable Fermi level in TCI SnTe.

In this study, we develop a chemical vapor deposition method based on Sn self-compensation to control defects and tune the Fermi level in SnTe. To do this, we induce Sn-rich conditions in our synthesis process by adding extra Sn precursors mixed with the SnTe precursor. X-ray photoelectron Spectroscopy (XPS) and Ultraviolet photoelectron spectroscopy (UPS) are combined to characterize the influence of the amount of Sn in the precursor on Sn vacancy defects and Fermi level of SnTe. The results provide valuable insights about the relationship between the chemical composition and electronic properties of SnTe.

## 2. EXPERIMENTAL

### *Synthesis of SnTe*

SnTe crystals were synthesized via chemical vapor deposition in a single zone tube furnace. Powders of 0.1g SnTe supplied by ALB Materials with 99.999% purity were measured and placed into a quartz boat. The boat was placed in a quartz tube with a diameter of 2.5cm at the center of a single-zone furnace. At the distance of 10 cm downstream of the precursor, a minimum of 4 mica substrates with a dimension of 1cm x 1cm were placed. Ultrapure Argon (Ar) gas was then flushed through the horizontal quartz tube at the rate of 1000 standard cubic centimetres per minute (sccm) in multiple bursts (maximum of 5 times) to purge residual moisture and oxygen from the tube. After each burst, the tube was allowed to achieve a pressure of 40 mTorr. After the last purge, the base pressure for the growth was maintained at 2 Torr by passing Ar gas through the tube at the rate of 40 sccm. The furnace was then set to ramp up to the growth temperature of 600 ℃ at a rate of 20 ℃/min, then dwell at this temperature for 10 min. When the growth is completed, the furnace is opened, and the tube was naturally cooled down while the Ar carrier gas continue to flow. The setup for this procedure is shown in Figure 1. To systematically study the effect of additional Sn on the Fermi level of SnTe, we chose three representative combinations for the precursor: SnTe only, SnTe+0.05g Sn, and SnTe+0.1g Sn.

Atomic force microscopy topography images was taken using Oxford Asylum MFP3D origin equipment. XPS and UPS measurements were performed using the Thermofischer NEXSA G2 equipment. XPS spectra was taken with a low power Al K-Alpha (1486.6 eV) beam with a $20\mu m$ spot size. Peak fitting was done using XPS Peak4.1. For each compositional analysis, the crystals were sputtered to remove the oxidized top surfaces of the as-grown crystals with an EX06 monatomic ion source. The UPS spectra was taken with a He I (21.1eV) ultraviolet lamp and a -10V bias was applied to the samples in order to distinguish between the analyzer and the sample secondary cut-off.

## 3. Results and discussion

We name the samples made with only SnTe precursor, SnTe+0.05g Sn precursor, and SnTe+0.1g Sn precursor as S0, S1 and S2, respectively. To ensure long-term stability, samples are maintained either in a vacuum or within an inert atmosphere. This approach effectively preserves the compositional integrity of the samples. Should oxidation occur, the oxidized layer can be successfully removed using an Ar ion gun during XPS analysis (Figure S2 in the supplementary information).

Optical images of S0, S1 and S2 are shown in Figure 1b, 1c and 1d. On average, the thickness of the crytsals as obtained via AFM imaging is 1.2µm (refer to Figure S1 in supplementary information). From the optical images, cubic shaped crystals can be observed, correlating the expected face-centered cubic crystalline structure of SnTe with the (001) orientation[29,43,44]. Further, we can confirm that the shape of the crystals do not change for the synthesis with the Sn-rich condition (S0,S1 and S2).

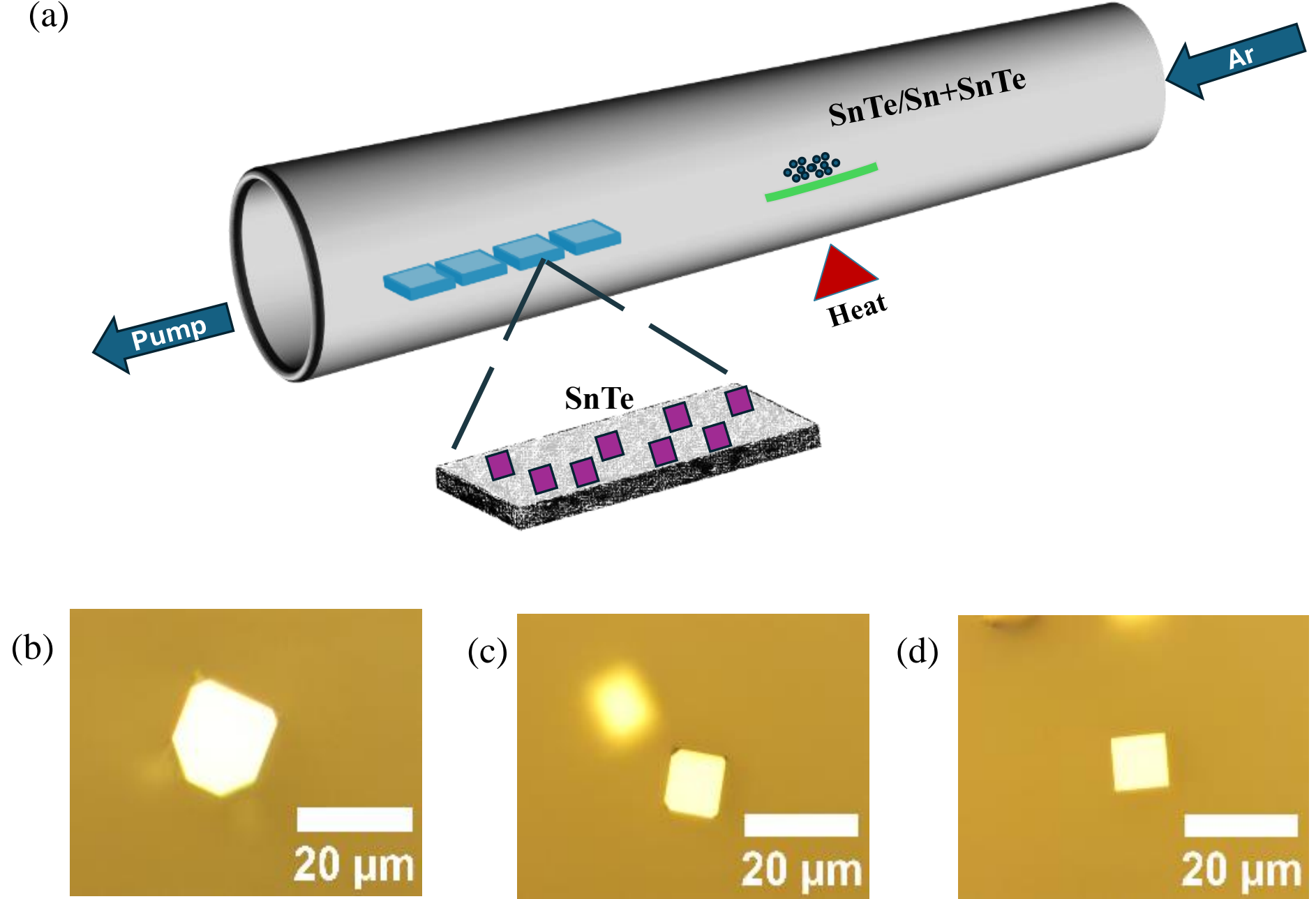

*Figure 1. (a) The schematic of the chemical vapour deposition setup for synthesizing SnTe. (b), (c) and (d) The optical microscope images of the as-grown SnTe crystals for samples S0, S1 and S2, respectively.*

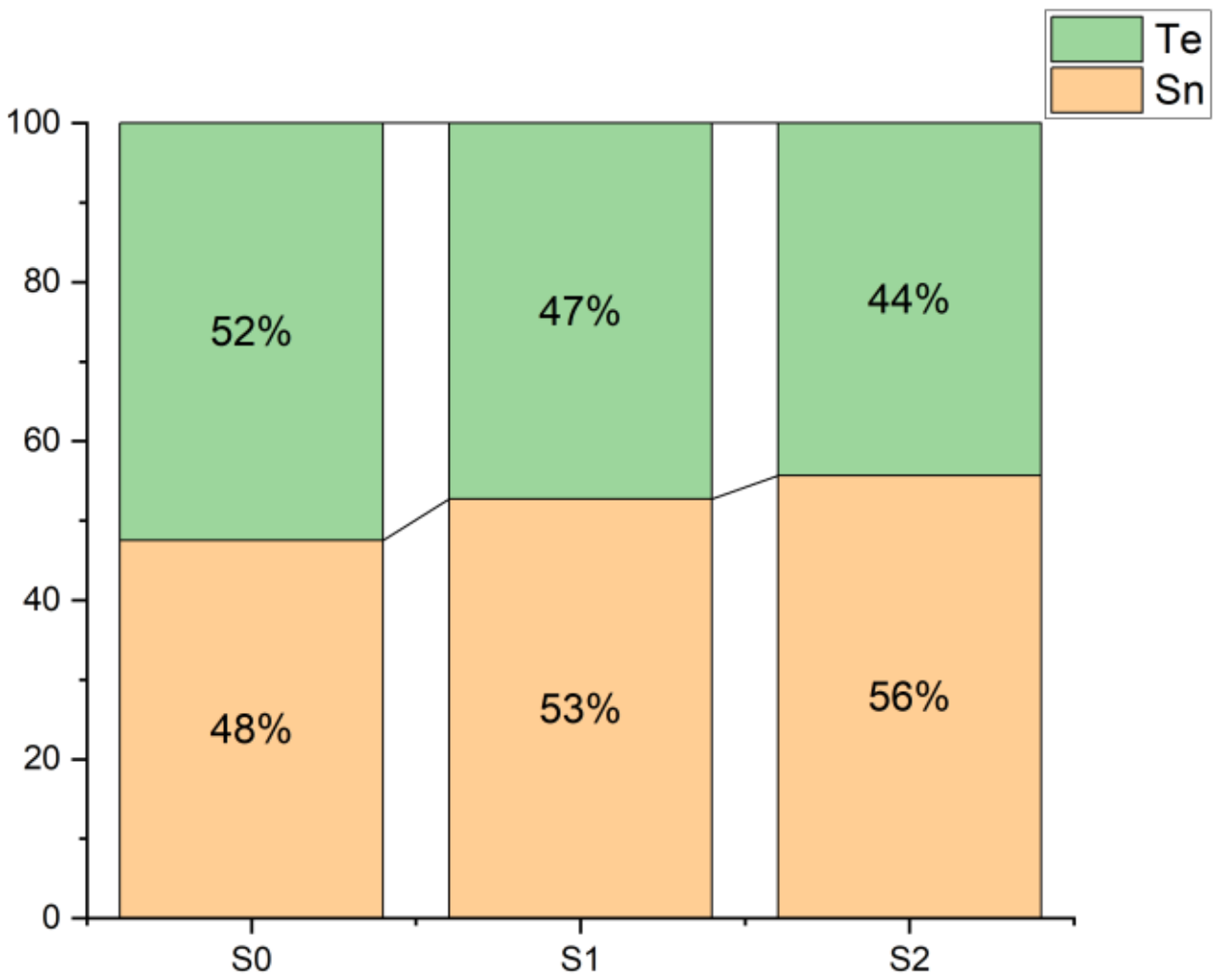

*Figure 2. Atomic percent composition of the as-grown crystals showing an increase in Sn composition with increase in Sn concentration in a Sn-rich synthesis environment.*

We first use XPS to investigate the chemical states of the Sn and Te elements in the S0 sample synthesized without adding Sn powders. The high-resolution core level peaks for Sn 3d and Te 3d orbitals are shown in Figure 3. Two peaks can be seen for both Sn and Te. The Sn $3d_{5/2}$ and $3d_{3/2}$ peaks are located at 484.92 eV and 493.34 eV, respectively. The separation of 8.4 eV between the peaks is consistent with existing literature values for the spin-orbit seperated energy gap[45,46]. The Te $3d_{5/2}$ and $3d_{3/2}$ peaks are located at 571.84 eV and 582.24 eV, with a spin-orbit spearation of 10.4 eV that is also consistent with existing literature[47]. The stoichiometric composition of the samples are determined by the areas of the fitted $Sn^{2+}$ 3d and $Te^{2-}$ 3d peaks of SnTe, as shown in Figure S3 in the Supplementary Information. For sample S2, the known[45,48–50] elemental Sn doublet peaks at 484.8 eV and 493.3 eV are detected and excluded from the stoichiometric ratio calculation. For the S0 sample, the Sn to Te ratio is calculated to be 48:52.

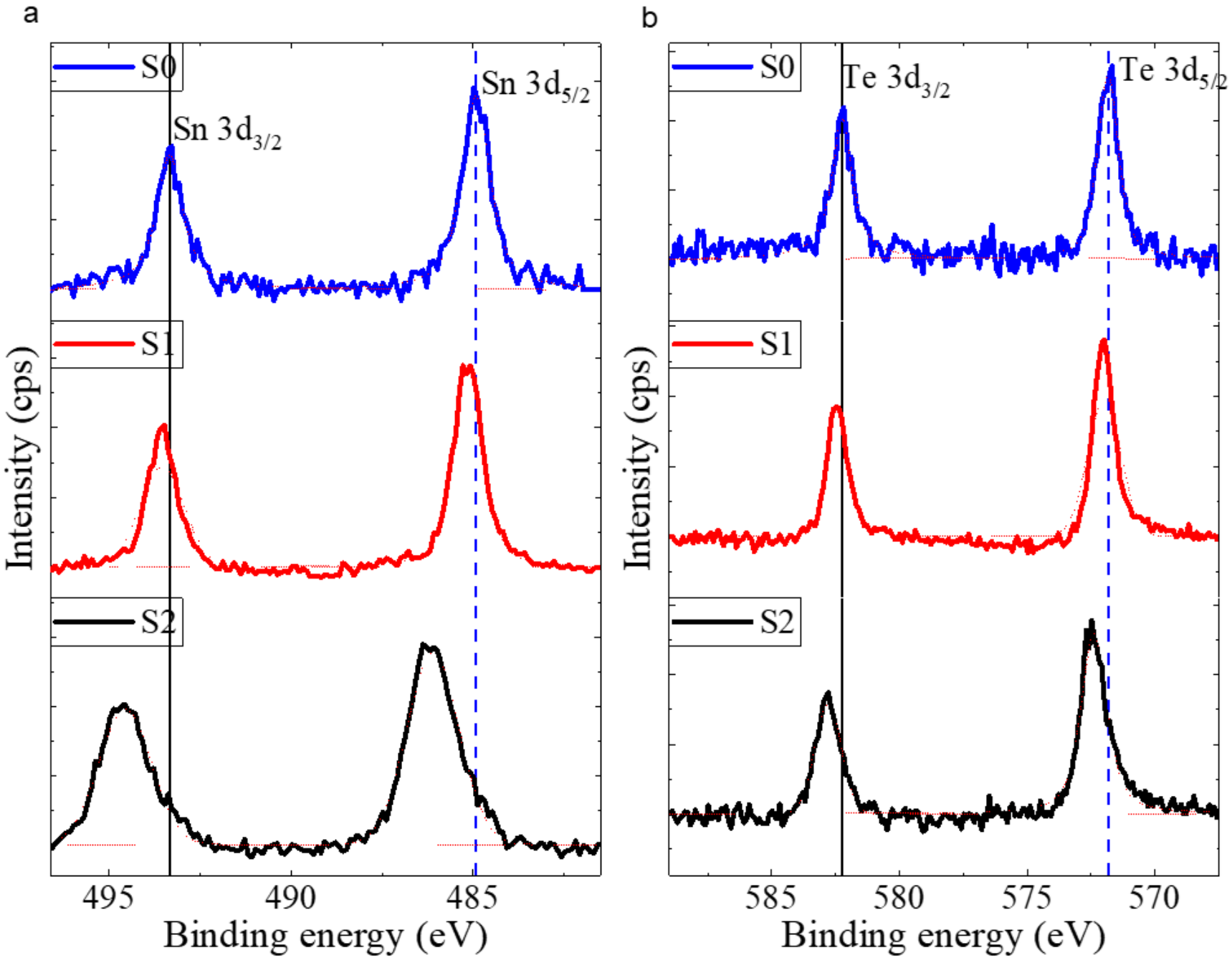

*Figure 3. High resolution 3d orbital scans of (a) Sn and (b) Te showing spin-orbit separated peaks. The blue broken lines serve as guide to see the peak shifts.*

For the S1 and S2 samples with added Sn powders in the precursor, XPS spectra show similar peaks, but with an overall shift in the binding energy. We first calculate the stoichiometric composition of these samples, and found that the percentage of Sn increases as the amount of Sn increases in the precursor. This is because a Sn-rich condition is created during the synthesis process with the additional Sn powders. This changes the formation energy of Sn vacancies and Te vacancies in SnTe as predicted by previous theoretical studies[18]. For sample S1, the Sn to Te ratio is 53:47, and for sample S2, the Sn to Te ratio is 56:44. The stoichiometric composition for samples S0, S1, and S2 are shown in Figure 2.

Next, we analyse the overall peak shift for samples S0, S1, and S2. The peak positions for the Sn and Te 3d peaks are shown in Table 1. Compared to sample S0, S1 and S2 show a blue shift in the Sn $3d_{5/2}$ peak with energies of 0.22 eV and 1.19 eV, and in the Sn $3d_{3/2}$ peak with energies of 485.14eV and 486.11eV respectively. As can be seen from this result, the blue shift in both the Sn $3d_{5/2}$ peak and Sn $3d_{3/2}$ peak show similar values, indicating a homogeneous shift as the amount of Sn increases in the precursor. The Te 3d peaks also show a similar homogeneous blue shift behaviour.

To understand the homogeneous peak shift, we look at how the kinetic energy of electrons measured in XPS is related to intrinsic properties of materials. In XPS, kinetic energy of electrons ejected from the material is measured through the photoelectric effect[38,51]:

$$E_{KE} = h\nu - E_B - \phi \qquad \ldots(1)$$

where $\nu$ is the incident radiation frequency, $E_B$ is the binding energy and $\phi$ is the work function of the material. The value of $\phi$ is determined by the Fermi level $E_F$ of the sample[52,53]:

$$\phi = E_{vac} - E_F \qquad \ldots(2)$$

The binding energy shown in XPS core-level spectroscopy is referenced to the Fermi level[54]. In terms of binding energy, equation 1 and 2 can be rewritten to show a direct relationship between the binding energy and Fermi energy as:

$$E_B = h\nu - E_{KE} - E_{vac} + E_F \qquad \ldots(3)$$

Equation 3 shows that the binding energy measured in XPS is closely related to the Fermi level in the sample. Thus, we attribute the homogeneous shift in the XPS core-level peaks to the change of Fermi level in SnTe. As the amount of Sn increases in the precursor, the Fermi level increases, causing a homogeneous blue shift in the XPS core-level peaks for both Sn and Te. This finding can be explained by the relationship between the Fermi level and defects established in semiconductors. SnTe naturally shows p-type conductivity at finite temperatures due to the presence of a large amount of Sn vacancies[23,55,56]. The Sn vacancies act as acceptors and create holes as charge carriers, which shift the Ferm level $E_F$ away from the middle of the band gap towards the valence band[57–59].

The relation between the density of carriers and the Fermi level can be described by the following equation[57,58]:

$$E_F = E_V - K_B T \ln \frac{N_A}{N_V} \qquad \ldots(4)$$

where $E_V$ is the energy at the top of the valence band, $N_A$ is the concentration of acceptors, while $N_V$ is the effective density of states in the valence band. Equation 4 shows that by modulating the density of acceptors, we can correspondingly tune the Fermi level of the material. As the concentration of acceptors reduces, the Fermi level increases in energy.

| Sample | Sn Precursor (g) | Sn | | Te | |
| --- | --- | --- | --- | --- | --- |
| | | $3d_{5/2}$ | $3d_{3/2}$ | $3d_{5/2}$ | $3d_{3/2}$ |
| **S0** | 0 | 484.92 | 493.34 | 571.84 | 582.24 |
| **S1** | 0.05 | 485.14 | 493.56 | 571.98 | 582.43 |
| **S2** | 0.1 | 486.11 | 494.56 | 572.39 | 582.79 |

*Table 1. Table showing the peak position of the Sn and Te 3d of the spin-orbit coupled peaks obtained from the high-resolution spectra scans.*

Combining this fact with equations 2 and 3, it can be deduced that the blue shift in binding energy from sample S0 to S2 is due to the reduction of Sn vacancy defects in SnTe, which increases the Fermi level. Compared to sample S0, the intrinsic SnTe, samples S1 and S2 are grown under a Sn-rich condition with additional Sn powders in the precursor. Therefore, our result shows that creating Sn-rich conditions in CVD is an effective approach to suppress Sn vacancies and tune the Fermi level in SnTe.

### 3.1 Workfunction by Ultraviolet Photoelectron Spectroscopy

Ultraviolet photoelectron spectroscopy (UPS) is employed to further study the electronic structure of the SnTe samples and quantify the Fermi level shift. UPS and its corresponding low binding energy measurements provides an important way to directly determine the work function and the Fermi level of materials[38–40]. By analysing the kinetic energy of valence electrons, precise measurements of the electrons' energy levels is provided, thus yielding an accurate determination of the Fermi level and work function of the materials[41,42]. A typical UPS spectra of our CVD synthesized SnTe sample is shown in Figure 4b, and the zoom-in sections in Figure 4a and Figure 4c are used to determine the work function and Fermi level.

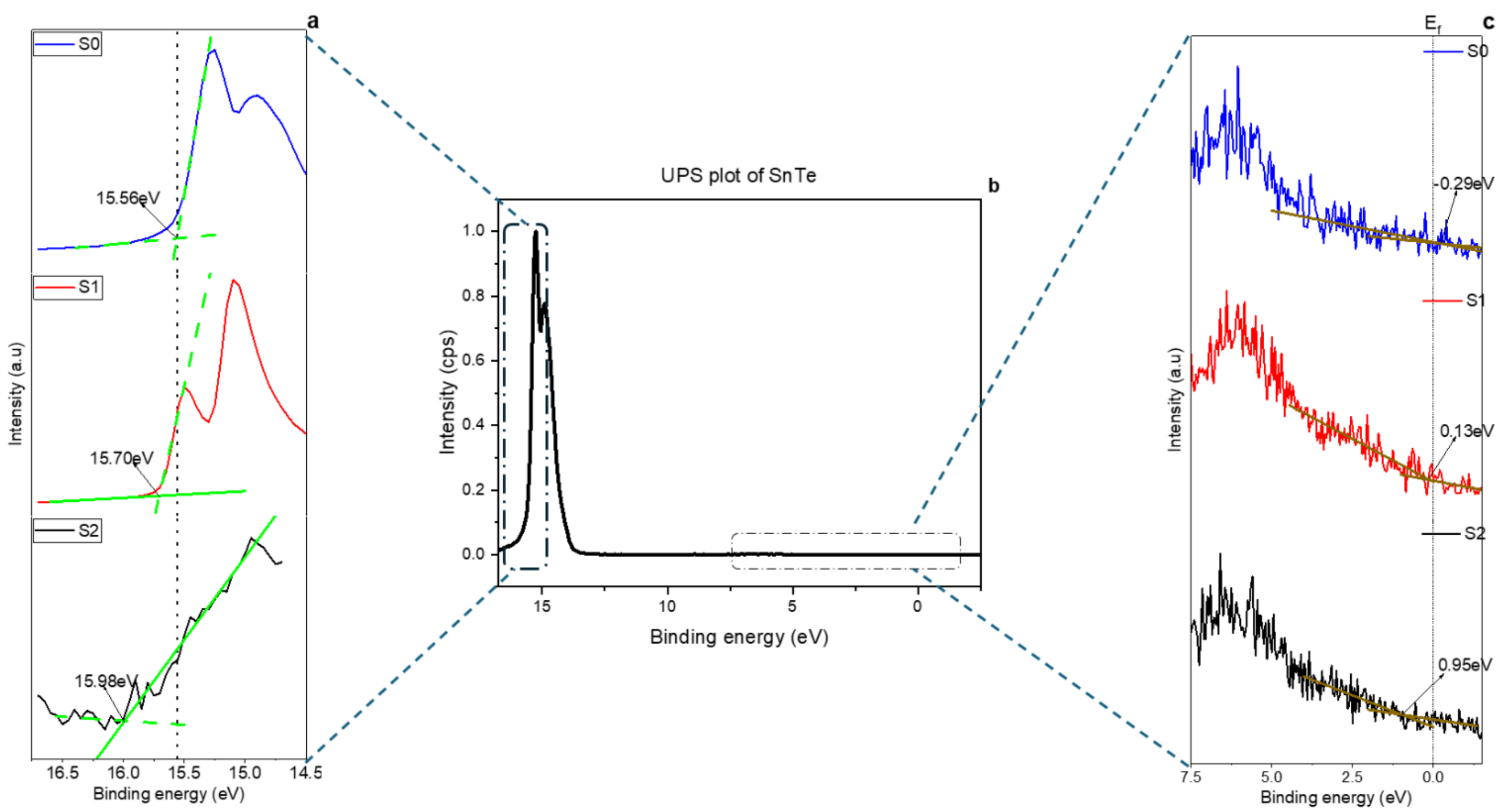

*Figure 4 UPS valence band spectra obtained from as-grown SnTe crystals with a bias of -10 V. (a) Zoomed out secondary electron cutoff region from which work function calculations were made; (b) Typical UPS spectra obtained for SnTe (c): Zoomed out expansion of the region near the Fermi level.*

To obtain the work-function, we determine the onset of the secondary electron cutoff (SECO) emission from the UPS spectra as shown in Figure 4a. Secondary electrons are the electrons that undergo inelastic scattering events within the material as they travel out of it during the UPS measurement[38,39]. These electrons are ejected out of the material with reduced kinetic energy. The UPS spectra is a result of contributions from the valence band electrons and these secondary electrons. Absolute values of work functions can be determined in UPS by measuring the minimum kinetic energy of secondary electrons generated by a known photon energy. This minimum is commonly referred to as the SECO as the low energy component is dominated by secondary electrons which typically provide a sharp cutoff in intensity. The SECO can be determined by linear fitting of the UPS spectra as shown in Figure 4a[39,42]. Thereafter, the workfunction can be obtained from the equation[39,50]:

$$\Phi = hv - \mathrm{SECO} \qquad \ldots(5)$$

where $hv$ here is 21.2eV, the radiation energy for the He I source impinging our crystals. By using the obtained values for SECO in Figure 4a, we can determine the work functions of our samples, and the result is shown in Table 2. Sample S0, the intrinsic SnTe, has the highest work function of 5.64 eV. With an increasing amount of Sn in the precursor for samples S1 and S2, the work function values are reduced to 5.5 eV and 5.22 eV. Since the work function is defined as the energy difference between the Fermi level and the vacuum level, a reduced work function indicates an increasing Fermi level, which agrees with the previous analysis on the peak shift of the core-level XPS spectra. Our finding agrees with an earlier report that shows the work function can be reduced by inreasing Sn content in SnTe polycrystalline films synthesized by magnetron sputtering[50].

| **Sample** | **SECO (eV)** | **Workfunction Φ (eV)** | **Fermi level ($E_F$) (eV)** |
|---|---|---|---|
| ***S0*** | 15.56 | 5.64 | -0.29 |
| ***S1*** | 15.7 | 5.5 | 0.13 |
| ***S2*** | 15.98 | 5.22 | 0.97 |

*Table 2 Calculated work function values for the as-grown crystals S0, S1 and S2. The Fermi level values obtained from the UPS spectra are also shown for the corresponding crystals.*

Furthermore, in order to confirm our finding, we quantitatively determine the position of the Fermi level for the as-synthesized SnTe samples. Using equation 3 and following previously reported methods[39,42,50,60,61], the relative position of the Fermi level with respect to intrinsic SnTe with no defects can be determined by linear fitting of the low binding energy region (near 0 eV) of the UPS spectra. Figure 4c shows the UPS plots in that region obtained from our samples. As seen from this high resolution valence band data, the Fermi level for our S0 sample was determned to be -0.29 eV, meaning it is 0.29 eV below the Fermi level of intrinsic SnTe with no defect, which is in the middle of the bulk bandgap. The negative value provides direct evidence of p-type doping caused by Sn vacancies[4,62–64], which aligns with our previous discussion. The data also show that with increasing Sn to Te ratio (S1 and S2), Fermi levels with positive values at 0.13 eV and 0.97 eV are observed, showing an upshift trend. The result agrees with our previous analysis on the work functions and peak shift in the core-level XPS spectra. The upward shift clearly indicates that creating Sn-rich conditions in the CVD process can successfully suppress Sn vacancies and raise the Fermi level in SnTe.

A notable observation in our results is that the estimated work function and the Fermi level do not vary to the same extent as predicted by equation 2. This discrepancy arises from the fact that the vacuum level $E_{vac}$ is not constant for all samples. While the $E_{vac}$ can be constant for metals, it is more complex for materials with a band gap, such as semiconductors and insulators[40]. In these materials, the vacuum level is influenced by the surface electrostatic potential, which depends on several factors. One key factor is the surface dipole, which results from the arrangement of atoms and molecules at the surface. Changes in surface composition, such as the adsorption of contaminants or the introduction of surface modifiers, can significantly alter the surface dipole, thereby shifting the vacuum level. Additionally, the surface morphology, including surface roughness and crystallographic orientation, plays a crucial role. Variations in surface roughness can lead to local electric field variations, which in turn affect the vacuum level. Similarly, different crystallographic orientations expose different atomic arrangements and densities, which can alter the surface dipole and thus the vacuum level[40,41,52].

In practice, surfaces are often inhomogeneous, leading to variations in the surface potential and in extension the vacuum level. In the case of SnTe, earlier works [50,65] have reported work function and Fermi level values in SnTe that do not vary to the same magnitude. Li *et al* [65] show results of DFT calculations signifying different vacuum levels $E_{vac}$ for pristine SnTe versus SnTe with tellurium defects. The different shifting magnitude of work function and Fermi level in SnTe is also observed in the experimental work by Yoon *et al*[50]. Therefore, we believe the different shifting magnitude observed in our work is due to the samples having different vacuum levels because of the different surface stoichiometry and defects.

## 4 Conclusion

In this study, we have demonstrated that the Fermi level in SnTe can be effectively tuned by manipulating the Sn-rich environment during the chemical vapor deposition (CVD) synthesis process. The described synthesis was repeated multiple times, and the method is highly reproducible in producing SnTe with tunable Fermi level. The method can be applied to the CVD synthesis of other material systems that utilize metal chalcogenides as precursors[66,67].

Our findings indicate that increasing the amount of Sn precursor during synthesis results in a blue shift in the high-resolution 3d XPS peaks for both Sn and Te, signifying an upward shift in the Fermi level. This shift is corroborated by quantitative analysis of the work function and Fermi level using UPS measurements. In future works, electronic transport measurements can be performed on our samples in order to study the effect of Fermi level on the transport behaviour of SnTe. Angle resolved photoemission spectroscopy can also be carried out to directly observe the band structure of our samples in *k*-space.

Our work presents an effective approach to achieve tunable Fermi levels in the topological crystalline insulator SnTe synthesized by a low-cost and scalable CVD method. It has the potential to optimize the electronic properties of SnTe for various technological applications including achieving faster and more efficient spintronics devices, realization of fault tolerant qubits necessary for stable and reliable quantum computing, highly efficient transistors and sensors among others. In these application areas, precise control over the Fermi level is necessary to access the topological surface states and harness their unique properties.

## Acknowledgements

Work by Bamidele Onipede and Hui Cai was supported by the startup fund at UC Merced and the Laboratory Directed Research and Development Program of Lawrence Berkeley National Laboratory under U.S. Department of Energy Contract No. DE-AC02-05CH11231. Work by Nisha Fletcher is supported by the National Science Foundation under Grant No. 2150531. The authors are sincerely grateful for the service provided by the IMF facility at UC Merced.

## Data availability statement

The data that support the findings of this study are available upon reasonable request from the authors.